# Classification of Complex Networks Based on Topological Properties

Burcu Kantarcı and Vincent Labatut
Computer Science Department
Galatasaray University
Istanbul, Turkey

*Abstract*—Complex networks are a powerful modeling tool, allowing the study of countless real-world systems. They have been used in very different domains such as computer science, biology, sociology, management, etc. Authors have been trying to characterize them using various measures such as degree distribution, transitivity or average distance. Their goal is to detect certain properties such as the small-world or scale-free properties. Previous works have shown some of these properties are present in many different systems, while others are characteristic of certain types of systems only. However, each one of these studies generally focuses on a very small number of topological measures and networks. In this work, we aim at using a more systematic approach. We first constitute a dataset of 152 publicly available networks, spanning over 7 different domains. We then process 14 different topological measures to characterize them in the most possible complete way. Finally, we apply standard data mining tools to analyze these data. A cluster analysis reveals it is possible to obtain two significantly distinct clusters of networks, corresponding roughly to a bisection of the domains modeled by the networks. On these data, the most discriminant measures are density, modularity, average degree and transitivity, and at a lesser extent, closeness and edgebetweenness centralities.

*Keywords—Complex Networks; Topological Measures, Properties Comparison, Cluster Analysis.*

## I. INTRODUCTION

A complex system is a specific type of real-world system, i.e. a set of interacting elements relatively isolated from their environment, and possessing some emerging properties [1]. Such a property is not present at the level of a single element, but appears when considering the system as a whole. Its study consequently requires focusing on the interactions between the system elements. For this purpose, graphs are a very appropriate modeling tool, in which elements and their relations are represented by nodes and links, respectively. And indeed, they have been used as such in a number of domains such as computer science, physics, biology, sociology, etc. [2]. The graph representation of a complex system is called a *complex network*. Such a graph has non-trivial topological properties, due to the specific features of the complex system it represents. Concretely, this means complex networks differ from both regular and random graphs.

Graphs can be characterized by many different measures, each one reflecting some particular traits of the studied structure. One can cite degree, transitivity, distance between nodes, density, etc. Some of these measures have been used to detect certain properties, seemingly very widespread in complex systems. For instance, it is now well known that many complex networks are scale-free, meaning their degree is power-law distributed [2]. Many of them also possess the small-world property, i.e. the average distance between their nodes increases only logarithmically with the number of nodes [3]. Complex networks are also known to have a transitivity several order of magnitude larger than that of random graphs of the same size [2]. It is also very common for complex networks to display a hierarchical or a community structure [2].

In the past, authors have focused on one or a few properties and studied them on networks representing a range of systems, with the purpose of showing their omnipresence. For example, in [3], Watts & Strogatz considered the transitivity and average distance in social, electrical and biomolecular networks, and found out they all behave similarly. On the contrary, other studies tried to show some properties are characteristic only of a certain class of networks. For example, in [4], Lancichinetti *et al.* observed different topological traits in community structures, depending on whether the considered data correspond to a biological, social, information, communication or computer network. These works highlight the importance of discovering regularities and discrepancies in complex networks topological properties. Indeed, these properties correspond to functional features. For example, a scale-free network is known to be sensitive to targeted attacks or failures, but resilient to random ones [1]. Topologically similar networks are therefore likely to represent systems with functional similarities, whereas network classes with specific topologic properties probably have unique functional features. However, existing works focused on a small number of networks and/or of properties. The network number limitation might be due to the difficulty of accessing data at this time. And regarding the focus on a few properties, this might be because those works were conducted to verify an *a priori* hypothesis. For example, one goal of Watts & Strogatz was to check if the small-world property was present also in non-social networks [3].

In this work, we propose to adopt a systematic approach in the study and comparison of the topological properties of complex networks. First, it is now possible to retrieve many publicly available network datasets through the Web, which allows considering a number of different systems. Second, data mining techniques are able to consider a large number of topological measures simultaneously, and to automatically identify the relevant ones. By considering many of them at the same time, we can find how they relate, which is not possible when focusing only on a few of them. We first gathered a dataset of networks spanning several domains, which constitutes our first contribution. We processed the most widespread topological measures for these networks, and used them as feature vectors to characterize them. We then applied standard data mining tools to study them depending on these features. We finally analyzed and interpreted this outcome.

The rest of the article is organized as follows. In the next section, we define the measures we selected to characterize the networks. In section III, we describe the tools we used to analyze the dataset. In section IV, we first present our dataset, before describing and interpreting our results. We conclude with a discussion of our work, its limitations and how these can be solved.

## II. TOPOLOGICAL MEASURES

In this section, we briefly describe the topological measures we used in our study. We decided to focus on the most popular ones in the network analysis literature. Here, we distinguish between local and global measures, i.e. those concerned with individual nodes or links, and those describing the network as a whole. In a given network, we note $V$ the node set and $E$ the link set. The number of nodes is $n = |V|$ and that of links is $m = |E|$. We note A the adjacency matrix, whose binary element $a_{uv}$ is 1 if there is a link between nodes $u$ and $v$, and 0 otherwise.

### A. Local Measures

**Degree.** This nodal measure $k(u)$ corresponds to the number of links attached to a node $u$.. In real-world networks, it often follows a power law, leading to the so-called *scale-free property* [2]. The degree $k(u)$ of a node $u$ can be formally defined using the adjacency matrix:

$$k(u) = \sum_{v \in V} a_{uv} \quad (1)$$

**Distance.** The *geodesic* distance $d(u,v)$ between two nodes $u$ and $v$ corresponds to the length of the shortest path between them. The distance distribution has been especially studied in the context of computer networks such as the Internet. When it increases logarithmically with the size $n$ of the network, the system has the so-called *small-world property* [2].

**Eccentricity.** This nodal measure $e(u)$ corresponds to the largest distance between a node $u$ and any other node $v \in V \setminus u$ [5].

$$e(u) = max_{v \in V} d(u,v) \quad (2)$$

**Betweenness centrality**. This nodal measure $c_B(u)$ is the number of shortest paths going through a node $u$. Let us note $\sigma_{vw}$ the total number of shortest paths between nodes $v$ and $w$, and $\sigma_{vw}(u)$ the number of shortest paths between $v$ and $w$ going through $u$ [5]. We have:

$$c_B(u) = \sum_{v<w \neq u} \frac{\sigma_{vw}(u)}{\sigma_{vw}} \quad (3)$$

**Closeness centrality.** This nodal measure $c_C(u)$ is the inverse of the sum of distances between the node of interest $u$ and all the other nodes $v \in V \setminus u$. It quantifies how close a node is from the rest of the network, in average.

$$c_C(u) = \frac{1}{\sum_{v \neq u} d(u,v)} \quad (4)$$

**Edgebetweenness.** This measure $C_E(e)$ represents the number of shortest paths containing the link $e$. Links with high edgebetweenness centrality corresponds to bridge-like connectors between two parts of a network [6]. In the following formula, $\sigma_{uv}(e)$ is the number of shortest paths between $u$ and $v$ containing the link $e$.

$$C_E(e) = \sum_{u<v} \frac{\sigma_{uv}(e)}{\sigma_{uv}} \quad (5)$$

**Local transitivity.** This nodal measure $C(u)$, also called *clustering coefficient*, corresponds to a ratio: the number of triangles including the node of interest $u$, to the number of possible triangles centered on this node. It can be interpreted as the probability for a link to exist between two randomly picked neighbors of the node of interest [3]. Let us note $N(u)$ the neighborhood of node $u$, then we have:

$$C(u) = \frac{|e_{vw} \in E: v,w \in N(u)|}{k_i(k_i - 1)/2} \quad (6)$$

### B. Global Measures

**Density.** This global measure noted $\delta(G)$ corresponds to the ratio of existing to possible links in network $G$. It ranges from 0 (no link at all) to 1 (all nodes are connected). Real-world networks are generally considered to be very sparse, with densities close to 0.1.

$$\delta(G) = \frac{m}{n(n-1)} \quad (7)$$

**Diameter & radius.** The diameter $D(G)$ is the maximum eccentricity over the network, i.e. the maximal distance between two nodes in $G$:

$$D(G) = \max_{u \in G} e(u) \quad (8)$$

On the contrary, the radius $R(G)$ is the minimum eccentricity of the network [7], i.e.:

$$R(G) = \min_{u \in G} e(u) \quad (9)$$

**Transitivity.** This measure, also called *clustering coefficient*, corresponds to the proportion of triangles in the network [2]. As such, it ranges from 0 (no triangles) to 1 (all possible triangles exist). It can be interpreted, when picking randomly a node, as the probability for two of its neighbors to be connected. According to the literature, the transitivity ranges from 0.1 to 0.8 in real-world network [2]. Let us note $\gamma(G)$ the number of subgraph with 3 links and 3 nodes (i.e. triangles) and $\tau(G)$ is the number of subgraphs with *at least* 2 links and 3 nodes (i.e. triangles and incomplete triangles). Then, the global transitivity is:

$$C(G) = \frac{\gamma(G)}{\tau(G)} \quad (10)$$

**Modularity.** This measure assesses the quality of a community structure. It corresponds to the proportion of links located inside the communities, minus an estimation of the same quantity obtained for a null model. Consequently, its upper bound is 1 while 0 means the community structure is equivalent to a random one. Values observed in real-world networks possessing a community structure are relatively high, in general. Let us define the function $\delta$ such that $\delta(u,v)$ is equal to 1 if nodes $u$ and $v$ belong to the same community, and 0 otherwise [8]. Then the modularity is:

$$Q = \frac{1}{2m} \sum_{u,v \in V} \left[ A_{uv} - \frac{k_u k_v}{2m} \right] \delta(u,v) \quad (11)$$

We note $Q(G)$ an estimation of the maximal modularity one can get on network $G$.

**Averages.** Besides the mentioned measures, which are global by construction, we also consider as global measures the averages of the previously listed local measures: average distance, average local transitivity, etc.

### III. ANALYSIS METHODS

In this section, we describe the methods we applied to analyze our dataset. We first explain the method we used to normalize the data, in order to be able to compare the various considered measures. We then describe the clustering tools we selected to detect groups of networks. Finally, we present the measures we used to assess the cluster qualities and compare them.

#### A. Preprocessing

The first step in our preprocessing is to normalize the values obtained for the selected measures. Then, these normalized data can be used to process the distance between networks.

**Normalization.** Not all measures are defined on the same range. This can be a problem when performing cluster analysis, because certain algorithms are sensitive to large differences in attribute values, and might give more importance to those with the largest magnitude [9]. To avoid this, we apply the same *Min-Max* normalization to all measures. Let $X = (x_1, \ldots, x_k)$ be a vector of values we want to normalize, where $k$ is here the number of values. Then, the Min-Max normalization consists in processing $X' = (x_1', \ldots, x_k')$ such as:

$$x_i' = \frac{x_i - \min_{1 \leq j \leq k} x_j}{\max_{1 \leq j \leq k} x_j - \min_{1 \leq j \leq k} x_j} \quad (12)$$

However, we treat local and global measures differently. For the global ones, each network is described by a single value: we consequently normalize a series of values over the dataset. This way, we can consistently compare networks, knowing the minimal and maximal values a network can reach in our dataset are 0 and 1, respectively.

For the local measures, we obtain a series of values for each network. Each value corresponds to a node, a link or a pair of nodes. We separately normalize these series, so that 0 and 1 now correspond to the minimal and maximal values one can find for a given network (and not for the whole dataset, like before). Indeed, our goal here is to compare distributions, so relative differences are more important than absolute ones. After this normalization, the distribution is processed under the form of histograms, containing 20 bins with a step of 0.05.

**Distance Matrix.** Once the measures have been normalized, it is possible for us to process the distances. We first calculate a partial distance for each measure taken separately. We then combine all of them to get the overall distance between two networks. Again, we distinguish between global and local measures.

For each global measure, we compare two networks by simply considering the *Manhattan* distance $d_M$, i.e. the absolute value of the differences. The result is necessarily a value ranging from 0 to 1, thanks to our normalization. For the local measures, we process the *Earth Mover's* (EM) distance $d_{EM}$ [10] between both histograms. Because we apply the EM distance to normalized histograms, it ranges from 0 to 1.

The overall distance is obtained by simply averaging all partial distances. By construction, it will also range from 0 to 1. The overall distance is processed for each pair of networks, in order to build the matrix distance required by the clustering algorithms. Let us note $n$ the number of global measures, and $m$ the number of local measures. Each normalized global measure is noted $M_i$ ($1 \leq i \leq n$) and each normalized local measure is noted $N_j$ ($1 \leq j \leq m$). The overall distance $d_O$ between two networks $G$ and $H$ is:

$$d_O(G, H) = \frac{1}{n} \sum_{1 \leq i \leq n} d_M(M_i(G), M_i(H)) \\ + \frac{1}{m} \sum_{1 \leq j \leq m} d_{EM}(N_j(G), N_j(H)) \quad (13)$$

#### B. Cluster Detection

Cluster analysis consists in empirically forming groups of objects, called clusters, with high intra-cluster similarity and low inter-cluster similarity. One can distinguish various general approaches: partitional, hierarchical and density-based algorithms. We chose to apply one tool of each these families, in order to be able to compare them and obtain consensual clusters.

**Partitional approaches**. They first split the dataset in several mutually exclusive clusters, and then maximize (resp. minimize) the intra-cluster (resp. inter-cluster) similarity by moving objects from one cluster to another. The most widespread algorithm is $k$-means, however it requires to perform averages objects, which we can since we only have access to the distances. For this reason, we selected a variant of $k$-means called Pam (*Partitioning Around Medoids*) [11]. This approach requires the user to specify the number of desired clusters. When this information is unknown, as it is for us, the classic approach consists in using a broad range of values, and selecting the one leading to the best clusters according to some criterion.

**Hierarchical approaches.** They build a hierarchy of clusters, called *dendrogram*. Two different methods exist for this matter: bottom-up or agglomerative, and top-down or divisive. In the former, each object is initially considered as a cluster, and those are iteratively merged until only one cluster containing all objects remains. In the latter, on the contrary, all the objects are in the same unique cluster, which is then repeatedly divided until obtaining only singleton clusters. The choice of the final clusters is made by selecting a level, called cut, in the dendrogram, according to some criterion of interest. We selected two popular algorithms: one agglomerative tool, Agnes (*Agglomerative Nesting*) [12], and one divisive tool, Diana (*Divisive Analysis*) [12].

**Density-based approaches.** Starting from an initial object called seed; a cluster is constituted by iteratively aggregating close objects. The cluster grows as long as some conditions regarding its density still hold. When it is not the case anymore, the cluster is complete (its limits have been reached). Another seed is then picked, in order to constitute a new cluster. We selected the DBscan implementation of this approach [13]. It requires the user to specify two parameters allowing defining the notion of cluster density: first, a radius defining the neighborhood of an object; second, the minimal number of object required inside this neighborhood, so that it is considered as dense.

*C. Cluster Evaluation*

Each selected clustering method outputs several partitions of the dataset, either because it is hierarchical (Diana, Agnes) or because some parameters must be tuned (Pam, DBscan). In order to identify the best partitions, we therefore need to be able to quantify their quality. For this matter, we used the average Silhouette width. Moreover, once the best partition has been identified for each tool, we want to compare them from tool to tool, in order to check for agreement. For this purpose, we used the Adjusted Rand Index.

**Silhouette.** This measure is based on two quantities noted $a(i)$ and $b(i)$ [14]. The former is the *average distance* between an object of interest $i$ and the rest of the objects located in the *same cluster*. For the latter, we first perform the same operation, but for objects located in a different cluster. $b(i)$ is the minimum of this quantity processed over all other clusters. From $a(i)$ and $b(i)$, we can process the silhouette width for the object of interest:

$$s(i) = \frac{b(i) - a(i)}{\max(a(i), b(i))} \quad (14)$$

The overall value is obtained by averaging $s(i)$ over all objects. Its range is $[-1; 1]$, and higher value means better quality.

**Adjusted Rand index.** This measure was designed to compare two partitions of the same set [15]. Let us note $n_{ij}$ the number of instances belonging to cluster $i$ in the first partition, and to cluster $j$ in the second one. We can then note $n_i$ the number of instances belonging to cluster $i$ in the first partition, whatever their cluster in the second partition is, and $n_j$ the symmetric quantity: number of instances in cluster $j$, independently from their cluster in the first partition. The Adjusted Rand Index is defined as:

$$ARI = \frac{\sum_{i,j}\binom{n_{ij}}{2} - [\sum_i\binom{n_i}{2}\sum_j\binom{n_j}{2}]/\binom{n}{2}}{\frac{1}{2}[\sum_i\binom{n_i}{2} + \sum_j\binom{n_j}{2}] - [\sum_i\binom{n_i}{2}\sum_j\binom{n_j}{2}]/\binom{n}{2}} \quad (15)$$

A value of 1 corresponds to perfect agreement, while 0 means random agreement.

IV. RESULTS AND DISCUSSION

Our dataset consists of a collection of 152 networks, all of them publicly available on the Internet. One of our goals was to study how the type of system represented by the network affects its topology. For this reason, we grouped them in 7 different domains: social interactions, scientific citations, communication, ecological systems, biomolecular interactions, computer networks and transportation systems. Social networks correspond to acquaintances, sexual and trust networks. Scientific citations represent bibliographic references. Communication networks include email and phone networks. Ecological networks are constituted of taxa and their predator-prey relationships. Biomolecular networks include protein, metabolic and genetic interaction networks. Computer networks include various representations of the Internet and the Web. Transportation networks correspond to airport interconnections and road systems. The number of networks for each domain is represented in TABLE I. In this section, we present our analysis of these data, using the methods described in section III.

TABLE I. DISTRIBUTION OF NETWORKS OVER DOMAINS

| Domain | Number of Networks |
|---|---|
| Social | 25 |
| Citation | 20 |
| Communication | 28 |
| Ecology | 20 |
| Biomolecular | 32 |
| Computer | 21 |
| Transportation | 5 |

*A. Topological Properties*

Let us describe our datasets in terms of the topological measures presented in section II. Those results are summarized in TABLE II, which contains lower and upper bounds, mean ($\mu$) and standard deviations ($\sigma$) for the main measures, and for each domain. The notations are the same than in section II, namely: size, density, average degree, global transitivity, average distance, diameter, radius and modularity.

**Size.** For all domains, the size of the smallest networks is of the same order of magnitude: a few tens of nodes. However, this is not the case for the largest ones. The largest Ecological and Transportation networks contain a few hundred nodes. For Social, Communication and Biomolecular networks, it is several thousand nodes. And Citation and Computer Science networks reach several tens of thousands of nodes. This highlights the fact real-world network sizes are very heterogeneous, spanning 3 orders of magnitude. This is confirmed by the generally large standard deviations.

**Density.** Similarly to what can be observed in the literature, most of our networks are very sparse, as seen in the average density and standard deviation of all domains. For some of them, the density is even as low as $10^{-4}$. However, the average density of Social and Transportation networks is clearly higher (roughly the double of the others). Moreover, some networks are remarkably dense in the Social, Communication and Biomolecular domains, as highlighted by their upper bounds.

**Degree.** According to the Kolomogorov-Simirnov tests we performed, all the studied networks have a power-law-distributed degree, a prominent feature in complex networks literature. For most domains, degree bounds have the same order of magnitude: a few units for the lower bound, several tens for the upper bound. The exceptions are Transportation, Communication and Citation networks, whose upper bounds reach several hundreds. For the Citation domain, this can be explained by the fact the networks are larger (in terms of nodes), compared to other domains, while they are as dense. For the Transportation and Communication domains, the networks are small but dense, which can explain these high upper bounds.

**Transitivity.** The literature highlights the fact real-world networks generally have a high transitivity. It does not seem to be the case so much when looking at the average values obtained on our dataset, which range from 0.07 to 0.40. A look at the bounds shows us the smallest values are almost zero, and the highest ones are not so large (around $0.5 - 0.6$), with the exception of Social and Transportation networks (0.86 and 0.84, respectively). The relatively large standard deviations highlight the heterogeneity of the networks in terms of transitivity. However, when comparing with values expected for ER networks with the same size and density, it turns out the networks of our dataset are more transitive.

**Distance.** The order of magnitude of the average distance and both distance bounds are roughly the same for all domains: the lower bounds are close to 1, the upper bounds are close to 10, and the average distances lie in between. All networks consequently have a very small average distance, when compared to their size in terms of nodes. Larger networks have a higher distance, but the increase is marginal. The observed average distances are higher than those expected for ER random networks of same size and density. This means the observed values alone are not sufficient to decide if the networks are small-world.

**Eccentricity.** For most networks, we observe a bimodal distribution of eccentricity, most of the nodes having very low or very high values. In terms of diameter, the order of magnitude of the diameter is the same for most domains, independently from the network size: it ranges from a few hops to a few tens. However, this is not true for the Social and Ecological networks, since the upper bound is tens of thousands of hops for them. This means that, even if the average distance is of the same order of magnitude than in other domains, it is possible for nodes to be much farther from the network center in Social and Ecological networks. Interestingly, the same observation does not hold for the radius, which is roughly similar for most domains. Computer networks stand out though, with a radius of hundreds of hops, instead of tens for the other domains.

**Centrality.** For most networks, the betweenness and edgebetweenness centralities are homogeneous, following a normal-like distribution. This means that, in a given network, most nodes and links lie on the same number of shortest paths, and only a few have extreme values. The presence of only a few central links supports the assumption the networks are modular: those links are known to connect communities. On the contrary, the closeness centrality distribution is bimodal. Both modes are extreme values like for the eccentricity.

**Modularity.** The modularity ranges from close to zero, or even slightly negative values, to as high as 0.93. Most networks have a clearly non-zero modularity, though. The most modular networks belong to the Citation domain. Most domains have a relatively high average ($0.3 - 0.4$). However, this is not the case of the Transportation and Ecological domains, whose average modularity values are 0.15 for the former and almost zero for the later. Thus, modularity seems to be exceptional for those domains, whereas it is the norm for the other ones.

TABLE II. OVERVIEW OF TOPOLOGICAL MEASURES RELATIVELY TO DOMAINS

|  | $n$ | $\delta$ | $\langle k \rangle$ | $C$ | $\langle d \rangle$ | $D$ | $R$ | $Q$ |
|---|---|---|---|---|---|---|---|---|
| **Social** | [11, 1882]<br>$\mu$:143.88<br>$\sigma$: 448.52 | [0.0004, 0.38]<br>$\mu$: 0,29<br>$\sigma$: 0,25 | [1.85, 66.69]<br>$\mu$: 11.39<br>$\sigma$: 14.54 | [0.01, 0.87]<br>$\mu$: 0.38<br>$\sigma$:0.25 | [1.26, 9.33]<br>$\mu$: 2.80<br>$\sigma$: 1.68 | [2, 305124]<br>$\mu$: 12212.12<br>$\sigma$: 61023.31 | [2, 16]<br>$\mu$: 3.2<br>$\sigma$:4.07 | [-0,03, 0.89]<br>$\mu$: 0.31<br>$\sigma$: 0.29 |
| **Citation** | [35, 27779]<br>$\mu$:3424.53<br>$\sigma$: 7547.97 | [0.0004, 0.26]<br>$\mu$: 0.07<br>$\sigma$: 0.09 | [3.24, 516.80]<br>$\mu$: 39.81<br>$\sigma$: 104.77 | [0.03, 0.69]<br>$\mu$: 0.23<br>$\sigma$: 0.17 | [1.76, 8.46]<br>$\mu$: 3.88<br>$\sigma$: 1.55 | [3, 37]<br>$\mu$: 13.93<br>$\sigma$: 0.26 | [2, 49]<br>$\mu$: 8.29<br>$\sigma$: 13.67 | [0.14, 0.93]<br>$\mu$: 0.41<br>$\sigma$: 0.20 |
| **Communication** | [12, 3861]<br>$\mu$: 427.93<br>$\sigma$:103.822 | [0.0004, 0.36]<br>$\mu$: 0.12<br>$\sigma$: 0.11 | [1.83, 27.70]<br>$\mu$: 7.50<br>$\sigma$: 5.66 | [0.01, 0.88]<br>$\mu$: 0.25<br>$\sigma$: 0.22 | [1.21, 6.53]<br>$\mu$: 2.98<br>$\sigma$: 1.50 | [3, 33]<br>$\mu$: 10.35<br>$\sigma$: 8.42 | [2, 22]<br>$\mu$: 5.25<br>$\sigma$: 6.64 | [0.01, 0.79]<br>$\mu$: 0.42<br>$\sigma$: 0.24 |
| **Ecological** | [24, 128]<br>$\mu$: 65.38<br>$\sigma$: 35.00 | [0.0816, 0.23]<br>$\mu$: 0.15<br>$\sigma$: 0.03 | [5.13, 33.39]<br>$\mu$: 18.15<br>$\sigma$: 10.11 | [0.25, 0.49]<br>$\mu$: 0.38<br>$\sigma$: 0.08 | [1.81, 3.36]<br>$\mu$: 2.31<br>$\sigma$: 0.35 | [8, 947493]<br>$\mu$: 133126.5<br>$\sigma$: 302590.7 | [2, 11]<br>$\mu$: 3<br>$\sigma$: 2.16 | [0.01, 0.53]<br>$\mu$: 0.04<br>$\sigma$: 0.12 |
| **Biomolecular** | [23, 3839]<br>$\mu$ 1099.44<br>$\sigma$:889.27 | [0.0012, 0.34]<br>$\mu$: 0.02<br>$\sigma$: 0.06 | [2.15, 15.88]<br>$\mu$: 5.34<br>$\sigma$: 2.37 | [0.02, 0.57]<br>$\mu$: 0.07<br>$\sigma$: 0.14 | [1.80, 7.65]<br>$\mu$: 4.66<br>$\sigma$: 1.16 | [3, 35]<br>$\mu$: 13.03<br>$\sigma$: 5.33 | [2, 63]<br>$\mu$: 9.79<br>$\sigma$: 15.90 | [0.01, 0.78]<br>$\mu$: 0.52<br>$\sigma$: 0.17 |
| **Computer** | [18, 10680]<br>$\mu$: 158.28<br>$\sigma$:2973.78 | [0.0002, 0.50]<br>$\mu$: 0.05<br>$\sigma$: 0.11 | [2.54, 39.1]<br>$\mu$: 6.95<br>$\sigma$: 8.67 | [0.01, 0.50]<br>$\mu$: 0.12<br>$\sigma$: 0.14 | [1.49, 18.98]<br>$\mu$: 4.31<br>$\sigma$: 3.48 | [2, 46]<br>$\mu$: 11.65<br>$\sigma$: 8.71 | [2, 352]<br>$\mu$: 38.13<br>$\sigma$: 86.11 | [0.01, 0.88]<br>$\mu$: 0.43<br>$\sigma$: 0.26 |
| **Transportation** | [75, 332]<br>$\mu$:174.40<br>$\sigma$: 107.60 | [0.0327, 0.24]<br>$\mu$: 0.22<br>$\sigma$: 0.26 | [4.23, 194,64]<br>$\mu$: 37.90<br>$\sigma$: 69.61 | [0.01, 0.84]<br>$\mu$: 0.32<br>$\sigma$: 0.26 | [1.21, 3.48]<br>$\mu$: 2.37<br>$\sigma$: 0.70 | [3, 19]<br>$\mu$: 6.94<br>$\sigma$: 6.27 | [2, 16]<br>$\mu$: 4.28<br>$\sigma$: 5.67 | [0.01, 0.44]<br>$\mu$: 0.15<br>$\sigma$: 0.16 |

*B. Correlation Study*

We now examine the correlations between the topological measures studied in the previous subsection. As mentioned before, we distinguish two types of measures: *global* and *local* ones. To ease the interpretation of our results, we split the correlation study in three parts: global vs. global, local vs. local and global vs. local.

**Global vs. global.** TABLE III shows the correlation between global measures only. Most of the values are close to zero, indicating no linear relationships between the measures. However, a few strong positive and negative correlations are also observed. The highest (0.76) one is measured between the density and transitivity, which can be explained by the fact that when a network becomes denser, the possibility to find triangles increases, too. The average distance and radius are also highly correlated (0.59). This is certainly due to the fact both measures reflect how compact the network is.

Density and transitivity are both negatively correlated to average distance (−0.45 and −0.43, respectively). When the network becomes denser, the average distance automatically decreases: because of the additional links, the shortest paths become even shorter. When the average distance is large, the probability for direct connections decreases, impacting the number of triangles.

Modularity is positively correlated with average distance (0.60), and like this measure, it is negatively correlated with both density and transitivity (−0.71 and −0.51 respectively). Indeed, the presence of a community structure requires links to be concentrated in communities. So, the network must be relatively sparse: if it is too dense, then the community structure cannot exist. The presence of a community structure increases the average distance: the sparsity of direct connections between nodes from different communities makes shortest paths longer, in average.

TABLE III. CORRELATION BETWEEN GLOBAL MEASURES

|   | $\delta$ | $\langle k \rangle$ | $C$ | $\langle d \rangle$ | $D$ | $R$ | $Q$ |
|---|---|---|---|---|---|---|---|
| $\delta$ | - | 0.16 | 0.76 | -0.45 | 0.02 | -0.14 | -0.71 |
| $\langle k \rangle$ | - | - | 0.12 | -0.16 | -0.01 | 0.00 | -0.13 |
| $C$ | - | - | - | -0.43 | 0.04 | -0.09 | -0.51 |
| $\langle d \rangle$ | - | - | - | - | -0.09 | 0.59 | 0.60 |
| $D$ | - | - | - | - | - | -0.03 | -0.12 |
| $R$ | - | - | - | - | - | - | 0.16 |
| $Q$ | - | - | - | - | - | - | - |

**Local vs. local.** Local measures take the form of distributions, so it is not possible to compare them directly using Pearson's coefficient. Instead, we considered two series constituted of the distances between these distributions, for each pair of networks in our dataset. So, we insist on the fact we do not consider the direct correlation between two measures here, but rather the correlation of the distances based on these measures. In other words, a strong correlation value means that if both measure are distributed similarly (resp. differently) in one network, they will also be distributed similarly (resp. differently) in the other. These results are presented in TABLE IV, using the same notation than in section II, namely: degree, local transitivity, eccentricity, betweenness, closeness and edgebetweenness.

Some measures are not correlated with any other: it is the case for edgebetweenness. On the contrary, we observe a relatively strong correlation between the remaining measures. This is particularly true of degree and local transitivity (1.00), which indicates their distributions change similarly from one network to another. This does not necessarily mean degree and transitivity are directly linearly dependent, but rather that when two networks have a similar degree distribution, they also have a similar transitivity distribution, and vice-versa. Betweenness centrality is strongly (eccentricity) or at least significantly (degree, transitivity, closeness) correlated with all other measures except edgebetweenness. More generally, the correlation between all measures but the edgebetweenness is never smaller than 0.23. This indicates there is a certain redundancy in the information conveyed by these measures.

TABLE IV. CORRELATION BETWEEN LOCAL MEASURE DISTANCES

|   | $k(u)$ | $C(u)$ | $e(u)$ | $C_B(u)$ | $C_C(u)$ | $C_{EB}(e)$ |
|---|---|---|---|---|---|---|
| $k(u)$ | - | 1.00 | 0.34 | 0.55 | 0.24 | 0.10 |
| $C(u)$ | - | - | 0.23 | 0.43 | 0.33 | 0.01 |
| $e(u)$ | - | - | - | 0.79 | 0.40 | -0.01 |
| $C_B(u)$ | - | - | - | - | 0.45 | 0.01 |
| $C_C(u)$ | - | - | - | - | - | -0.01 |
| $C_{EB}(e)$ | - | - | - | - | - | - |

**Global vs. local.** To study the correlation between global and local measures, we also used the distances. Here again, it is important to be cautious with our interpretation: a strong correlation means that when two networks are similar in terms of some global measure, they are also similar regarding the distribution of the considered local measure. TABLE V shows the results we obtained.

Most of the measures are not correlated. However, we observe a relatively strong positive correlation for some of them. The highest is observed for density and eccentricity (1.00). This means that, when two networks have the same density, they tend to have the same eccentricity distribution (and vice-versa). At a much lesser extent, the same remark can be made for the closeness and betweenness centrality.

TABLE V. CORRELATION BETWEEN GLOBAL AND LOCAL MEASURE DISTANCES

|   | $\delta$ | $D$ | $C$ | $Q$ | $\langle d \rangle$ | $\langle k \rangle$ | $R$ |
|---|---|---|---|---|---|---|---|
| $k(u)$ | 0.10 | -0.09 | 0.25 | 0.13 | 0.01 | 0.00 | 0.00 |
| $C_B(u)$ | 0.43 | -0.09 | 0.23 | 0.00 | 0.01 | 0.00 | 0.00 |
| $C_C(u)$ | 0.44 | 0.01 | 0.18 | 0.04 | -0.04 | 0.00 | 0.00 |
| $C(u)$ | 0.31 | -0.06 | 0.33 | 0.02 | 0.01 | 0.00 | 0.00 |
| $e(u)$ | 0.24 | -0.25 | 0.04 | -0.01 | -0.01 | 0.00 | 0.00 |
| $C_{EB}(e)$ | 1.00 | -0.12 | 0.43 | -0.01 | 0.07 | 0.00 | 0.00 |

Average degree, radius, average distance and modularity are correlated with no local measure. Average degree is not even correlated with degree distribution, and neither are radius and average distance with distance distribution. This observation is valuable, since it means those local measures are not summarized by the corresponding global measures, contrary to what one could *a priori* assume. For degree, this is consistent with our knowledge though: since it is power-laws-distributed, we know the average is not a characteristic value.

TABLE VI. SIGNIFICANT MEASURES FOR NETWORK DOMAINS

| | Biomolecular | Citation | Computer | Ecology | Transportation | Social | Communication |
|---|---|---|---|---|---|---|---|
| Biomolecular | - | $C$ | | $C, Q, \langle d \rangle, \delta$ | $C, Q, \langle d \rangle$ | $\delta\ C, Q, \langle d \rangle$ | $C, \langle d \rangle$ |
| Citation | | - | - | $\langle d \rangle$ | $\delta$ | $Q$ | $\delta$ |
| Computer | - | - | - | $C, Q, \langle d \rangle, \delta$ | $Q$ | $\delta, C, \langle d \rangle, Q$ | $\langle d \rangle$ |
| Ecology | - | - | - | - | | $Q$ | |
| Transportation | - | - | - | - | - | - | |
| Social | - | - | - | - | - | - | $\delta$ |
| Communication | - | - | - | - | - | - | - |

*C. Domain Comparison*

The domains constitute the natural partition of our dataset. In order to understand what makes them different, we performed an ANOVA. This analysis aims at identifying which global measures allow discriminating domains. The ANOVA reveals 4 measures are significantly different in at least one domain: average distance ($p < 10^{-3}$), density ($p < 10^{-6}$), modularity ($p < 10^{-3}$) and transitivity ($p < 10^{-6}$). We performed Tukey's post-hoc test to identify which domains have different average values for these measures. TABLE VI displays the significantly different measures by pair of domains.

All four measures are significantly different between Biomolecular and Computer networks on one side, and Ecology and Social networks on the other side. One the contrary, Biomolecular and Computer networks are not significantly different, and neither are Ecology and Social networks. Although it is not as marked, Transportation networks are also different from both Biomolecular and Computer networks, but not from Ecology and Social networks. Finally, Citation networks lie somewhere in between, since they differ in one measure from all domains but Computer.

In the end, a clear separation appears between two groups of domains. The first contains Biomolecular, Citation and Computer networks, and the other includes Ecology, Transportation, Social and Communication networks. The question is now to know if this separation, based on a subset of the global measures only, is confirmed when considering the whole available information, thanks to the cluster analysis.

*D. Network Clusters*

As mentioned in section III, we have applied all 4 selected clustering algorithms (Agnes, Diana, DBscan and PAM) over the whole dataset; using the Silhouette measure to identify the best partitions, and the Adjusted Rand Index (ARI) to compare them. All methods reach their maximal Silhouette value for 2 clusters. Diana has the highest Silhouette with 0.44, Pam being a close second with 0.42, followed by DBscan (0.40) and Agnes (0.39). These values are not very high (the Silhouette upper bound being 1), but they still show there is a non-random separation between two groups of networks, as the lower bound of Silhouette is −1.

The clusters found by Diana and Agnes have largely similar structures, with an ARI of 0.75. After them, Pam and Agnes show the second highest similarity with a 0.45 ARI, and Diana and PAM reach the value 0.41. On the contrary, the clusters found by DBscan are very different, since the ARI is almost zero when compared with all three other methods. Because of the nature of this algorithm, it certainly means it found non-convex clusters. Those are worth exploring, however in the rest of this work, we decided to focus on the clusters identified by Pam, because it is highly similar to both hierarchical algorithms, and it is very close to Diana in terms of Silhouette. Therefore, we aimed at making a trade-off between the cluster quality and agreement between algorithms.

TABLE VII represents the distribution of networks of different domains over the two clusters detected by Pam. While Biomolecular, Citation and Computer networks are largely grouped in the first cluster, Ecological, Transport, Social and Communication networks are mostly grouped in the second cluster. The first cluster is dominated by Biological networks, whereas Social and Communication clusters dominate the second one. Interestingly, these clusters confirm the partition we previously inferred from the ANOVA conducted over the domains, using the global measures alone. However, the bisection is finer, since it is performed at the level of networks, and not at that of the domains. This allows highlighting the fact a small minority networks do not have topological features typical of their own domain, and therefore constitute outliers worth studying in further details.

TABLE VII. DISTRIBUTION OF DOMAINS OVER CLUSTERS

| | Cluster 1 | Cluster 2 |
|---|---|---|
| Biomolecular | 29 | 3 |
| Citation | 16 | 4 |
| Computer | 19 | 2 |
| Ecology | 1 | 19 |
| Transportation | 0 | 5 |
| Social | 5 | 20 |
| Communication | 5 | 23 |

In order to identify the discriminant topological measures for our clusters, and compare them with those previously obtained for the domains, we conducted another ANOVA. It indicates not less than 9 measures differ significantly between the clusters. For global measures, we have transitivity ($p < 10^{-16}$), diameter ($p < 0.01$), modularity ($p < 10^{-16}$), average distance ($p < 10^{-8}$), density ($p < 10^{-9}$) and average degree ($p < 10^{-3}$). For local measures, it is closeness ($p < 10^{-3}$), local transitivity ($p < 10^{-9}$) and edgebetweenness ($p < 10^{-9}$).

Amongst the discriminant global measures, we find the 4 ones already identified when studying the domains: modularity,

transitivity, density and average distance. The additional global measures are diameter and average degree, which means the only global measures considered as not discriminant are the radius, average centralities and average eccentricity. Out of the 7 selected local measures, only the 3 mentioned above are considered discriminant, the remaining ones being betweenness, degree, distance and eccentricity. Interestingly, closeness and edgebetweenness are considered as not discriminant when averaged, but they are discriminant when handled as distributions. It would be interesting to investigate why averaging them results in an information loss.

## V. Conclusion

In recent works, specific topological properties have been identified as present in most complex networks, independently from the modeled system, or on the contrary, specific only to certain types of systems. In this work, we tried to extend this kind of work, by adopting a systematic approach. For this purpose, we constituted a dataset of 152 real-world complex networks, distributed over 7 applicative domains; and analyzed it using data mining tools. We first processed 14 widespread topological measures for each networks, including local and global ones. We then compared them relatively to the domains.

Some measures, such as density, average degree and transitivity are very heterogeneous, even when considering a single domain. On the contrary, other measures such as average distance and modularity are ether generally more homogeneous, or at least homogeneous when considering domains independently. A correlation study showed strong positive relations between certain global measures: density and transitivity, average distance and modularity; and a strong negative relation between density and modularity. In terms of local measures, there are strong relations between degree and transitivity, and between eccentricity and betweenness. On the contrary, edgebetweenness is not related to any other local measure, but it is strongly related to density.

An additional ANOVA performed on global measures, relatively to the domains, showed transitivity, modularity, average distance and density were the only significantly different measures. These differences are clear enough to allow distinguishing two groups of domains: Biomolecular, Citation and Computer networks on one side, and Ecology, Transportation, Social and Communication networks on the other. This dichotomy was confirmed by a cluster analysis based on both global and local measures, although it also allowed performing a finer separation of the networks, highlighting outliers amongst most domains. Moreover, the cluster analysis used additional topological measures to discriminate networks: diameter and average degree for global measures, and closeness centrality, local transitivity and edgebetweenness for local ones.

The first contribution of our work was to tackle the problem of identifying discriminant topological measures by using a systematic approach, in terms of amount of data, topological measures and analysis tools. By opposition, previous works focused only on one or two measures, and on a small number of networks and analysis tools. Our second contribution is the constitution of a real-world complex network dataset. Our third contribution is the analysis of this dataset, mainly based on correlation study, ANOVA and cluster analysis.

However, we are aware of certain limitations, too. First, the dataset is relatively small. We are currently still working on it, in order to considerably increase its size. Second, in this first work, we wanted to focus on the most widespread topological measures, but it is possible to consider many others, such as network centralization [5], fractal dimension [16], node roles [17], assortativity measures [18], etc. Third, more advanced tools can be used to study the relations between topological measures, and also to interpret the cluster analysis results. For example, discriminant analysis [19] would allow distinguishing topological measures in terms of discriminant power.